\documentclass[twocolumn,apl,amsmath,amssymb,showpacs,superscriptaddress]{revtex4-2}
\usepackage{epsf} 
\usepackage{graphicx}
\usepackage{color}
\usepackage{soul}
\usepackage{gensymb}
\usepackage{sidecap}
\usepackage{amsmath}
\usepackage{mathtools}
\usepackage{float}
 \usepackage{multirow}
\usepackage[hidelinks,colorlinks=true,linkcolor=blue,citecolor=blue]{hyperref}
\DeclareUnicodeCharacter{2212}{\textendash}
\usepackage{tabularx}

\begin{document}
\title{Improved Electrochemical Performance and Diﬀusion kinetics by Boron-doping in Na$_{0.66}$Mn$_{0.8}$Fe$_{0.2}$O$_{2}$ Layered Cathodes for Sodium-Ion Batteries}  
              
\author{Jayashree Pati}
\email{These authors contributed equally. }
\affiliation{Department of Physics, Indian Institute of Technology Delhi, Hauz Khas, New Delhi-110016, India}
\author{P. Senthilkumar}
\email{These authors contributed equally. }
\affiliation{Department of Physics, Indian Institute of Technology Delhi, Hauz Khas, New Delhi-110016, India}
\author{Deepak Seth}
\email{These authors contributed equally. }
\affiliation{Department of Chemical Engineering, Indian Institute of Technology Delhi, New Delhi 110016, India}
\author{Riya Gulati}
\email{These authors contributed equally. }
\affiliation{Department of Physics, Indian Institute of Technology Delhi, Hauz Khas, New Delhi-110016, India}
\email{These authors contributed equally. }
\author{Manish Kr. Singh}
\affiliation{Department of Physics, Indian Institute of Technology Delhi, Hauz Khas, New Delhi-110016, India}
\author{Madhav Sharma}
\affiliation{Department of Physics, Indian Institute of Technology Delhi, Hauz Khas, New Delhi-110016, India}
\author{Anita Dhaka}
\affiliation{Department of Physics, Hindu College, University of Delhi, New Delhi 110007,  India}
\author{M. Ali Haider}
\affiliation{Department of Chemical Engineering, Indian Institute of Technology Delhi, New Delhi 110016, India}
\affiliation{Indian Institute of Technology Delhi-Abu Dhabi, Khalifa City B, Abu Dhabi, UAE}
\author{Rajendra S. Dhaka}
\email{rsdhaka@physics.iitd.ac.in}
\affiliation{Department of Physics, Indian Institute of Technology Delhi, Hauz Khas, New Delhi-110016, India}

\date{\today}      

\begin{abstract}

We report the electrochemical investigation and study the diffusion kinetics of boron doped  Na$_{0.66}$Mn$_{0.8}$Fe$_{0.2}$O$_{2}$ (B-NMFO) cathode materials for sodium-ion batteries. Notably, the B-NMFO cathode exhibits improved specific capacity of 163 mAh g$^{-1}$ as compared to 133 mAhg$^{-1}$ at 0.1~C for the NMFO cathode. Further, we observe better capacity retention of 70\% for B-NMFO as compared to the NMFO (60\%) at 1 C after 200 cycles, indicating high structural stability due to the presence of strong B-O bonds. The diffusion coefficient evaluation through galvanostatic intermittent titration technique and cyclic voltammetry, which is found to be in the range of 10$^{-8}$--10$^{-10}$ cm$^{2}$s$^{-1}$. Interestingly, the temperature dependent distribution of relaxation time (DRT) analysis provides a clear understanding about the individual physical processes occurring at different time domains during the electro-chemical testing. Moreover, density functional theory is employed to determine the energetics and the electronic properties of B-NMFO, which suggests that the interstitial tetrahedral sites, especially those next to vacancies, are the dominant incorporation path ways for B in the host structure. Additionally, classical molecular dynamics (MD) simulations are applied to gain insights into the Na-ion transport properties in the bulk structures cathode materials. 
\end{abstract} 

\maketitle

\section{\noindent ~Introduction}

The sustainable energy resources with the growing economical and efficient conversion technologies are the decisive solution for the rising environmental issues and also paves a way towards the portable and uninterrupted sources of energy \cite{GoodenoughEES14}. In the era of increasing energy demand, the researchers are mostly investigating on the environment friendly sustainable energy sources, which are divided in three categories: (i) clean energy harvest/conversion through solar energy, wind, geo-thermal, mechanical and tidal energy; (ii) grid energy storage in form of chemical potential like batteries, fuel cells and hydrogen; (iii) lastly the smart and less energy consuming devices for efficient usage and management of energy \cite{ChenNC19, Sharma_CCR_25}. In the present time of modern world, the global call for environmental benign sustainable battery technology for energy storage steer its gear towards the EV (electric vehicle), portable devices (laptop, mobile etc.) industry, where chemical energy is converted to electrical energy. For the said purpose, lithium-ion batteries (LIB) have long been used owing to their high energy density and long cycling stability \cite{GoodenoughJACS13}. However, the natural scarcity of lithium in the earth’s crust and its high extraction cost cause a strong concern for its wide range applications. The parameters like battery cost, cycle life and energy/power density play vital roles in case of stationary applications, which require naturally abundant materials to produce cost-effective rechargeable battery technologies. With a global concern to solve these issues, sodium-ion batteries (SIBs) are considered as the efficient and cost-effective energy storage technology among the secondary rechargeable devices \cite{VaalmaNRM18}. Further, the SIBs deliver more safer and environmentally benign solutions to LIBs with almost similar electro-chemistry. But at the same time, the SIBs suffer from low gravimetric energy density (low specific capacity) and inferior cycle life as compared to the LIBs, irrespective of its potential use in high power stationary storage applications \cite{XuAEM22}. In this case, selection of potential cathode materials, play a vital role in determining the electro-chemical kinetics of SIBs, which must contain suitable lattice sites and interstitial spaces to accommodate and release active Na-ions, reversibly featuring high energy densities and prolonged cycle life \cite{LiangAEM23}. In the present date, three important categories of materials are being investigated as positive electrodes for SIBs: layered transition metal oxides, polyanionic compounds, and prussian blue analogs \cite{GaoESM20}. Here, each category of the cathode materials has their own features and inherent advantages and problems \cite{LiangAEM23}. 

The layered materials with large spacing for accommodating Na$^{+}$, have high reversible specific capacities, high energy densities, and excellent rate capabilities combined with susceptibly convertible technologies. However, such a layered structure is prone to collapse when cycling large-radius Na$^{+}$ (de)insertion, resulting in an unsatisfactory cycle lifespan; besides, most layered oxides are sensitive to the moisture in the air and the absorbent, thus bringing about storage difficulties \cite{HanEES15}. At the same time, layered oxides Na$_{x}$MO$_{2}$ (M = transition metals) are prominent cathode possibilities owing to their substantial specific capacity, straightforward production, and the versatility of metal ingredients. The series of Fe- and Mn-containing layered oxides Na$_{x}$Fe$_{1-y}$Mn$_{y}$O$_{2}$ have been widely investigated owing to their environmental friendliness, cost-effective metal components, and substantial reversible capacity \cite{GonzaloJMCA14, YabuuchiNM12}. Yabuuchi {\it et al.} studied the electrochemical performance of P2 and O3-type Na$_{2/3}$Fe$_{1/2}$Mn$_{1/2}$O$_{2}$ and observed a significant specific discharge capacity of 190 mAhg$^{-1}$ at 0.05~C for P2-type, whereas the O3-type delivers only a reversible specific capacity of 111 mAhg$^{-1}$ in a potential window of 1.5--4.3~V \cite{YabuuchiNM12}. The findings with varying stoichiometry inferred superior performance of P2-type as compared to O3-phase, and higher Mn concentration can yield increased capacity and  better structural stability \cite{GonzaloJMCA14, YabuuchiNM12}. Despite of high reversible capacities, these layered oxides suffer poor cycling stabilities owing to the Jahn-Teller distortion of Mn$^{3+}$, which affects the structural stability of Mn-based oxides during cycling \cite{CaballeroJMCA02}. In addition, the irreversible oxygen release from layered compounds accelerates electrolyte decomposition, causing a thick, resistive cathode electrolyte interphase (CEI) and slow Na$^{+}$ transfer kinetics \cite{PanJPS16}. A common strategy to mitigate these issues is to introduce foreign cations to reduce the Mn$^{3+}$ high spin concentration and stabilize the crystal structure \cite{YabuuchiJMCA14, CaoAEL19, MaJPS16}. 

In recent years, some investigations were carried out by substituting small amount of electrochemically inactive elements (Ti$^{4+}$, Al$^{3+}$, Nb$^{5+}$) in the Mn-rich P2 materials, which enhances the cycling stability by significantly increasing the oxygen binding energy to alleviate the irreversible oxygen redox reaction despite the low voltage profile of the Mn redox potential (low energy density) \cite{CaoAEL19, MaJPS16, LiuAEM18}. Also, these electrodes can be stabilized to endure prolonged cycling by substituting light element, which can have strong bond with oxygen and has the potential to mitigate the irreversible anionic redox and dissolution of Mn$^{3+}$ ions by maintaining a minimal capacity decrement \cite{XiaoJPCC12}. Interestingly, considering the facts, boron can be utilized as a dopant for the Mn-rich cathode materials as it has the highest oxygen binding energy (B--O = 809~kJmol$^{-1}$) than all the reported dopants \cite{LuoNY07}. The high binding energy of B with oxygen helps the system in tuning the lattice oxygen framework stabilizing structure, while enabling controlled oxygen non-stoichiometry enhance the redox activity having high reversible capacity \cite{PanJPS16, LiAFM14}. In a investigation of B-doped material, Li {\it et al.} reported the improved oxygen stability in Li-cathodes through the tuning of the electron density of O 2$p$ band by lowering the band level through boron doping \cite{LiCM17}. 

Therefore, in this work, we compare the electrochemical performance of Mn-rich Na$_{0.66}$Mn$_{0.8}$Fe$_{0.2}$O$_{2}$ (NMFO) and boron-doped NMFO (B-NMFO) cathode materials, synthesized through sol-gel process, to unveil the effect of boron doping strategy. We find B-NMFO cathode exhibits improved ($\approx$18\%) specific discharge capacity of 163~mAhg$^{-1}$ at 0.1~C with operating voltage of 2.7~V, as compared to NMFO, which delivers a reversible capacity of 133 ~mAhg$^{-1}$ in a voltage window of 1.5--4.2~V. The rate capability test of B-NMFO cathode confirms the sustainability upto high current rate of 5~C and retention of around 98\% when comes back to 0.2~C, signifies the highly reversible nature of the material. Intriguingly, the long cycling test at 1~C shows around 70\% capacity retention after 200 cycles. The bulk diffusivity of Na-ion was evaluated through GITT (galvanostatic intermittent titration technique) and CV (cyclic voltammetry), which falls in the range of 10$^{-8}$ to 10$^{-10}$ cm$^{2}$s$^{-1}$ for both the cathodes. Also, we used the DRT analysis for the B-NMFO cathode at different temperatures and analysed the individual electrochemical processes in the time domain. Moreover, density functional theory (DFT) is employed to determine the energetics and the electronic properties of boron-doped NMFO, which suggest that the B doping is site-selective, favouring BO$_{3}$--type configurations under Na-ion vacancies as indicated by the lower energy of formation of B-NMFO structure. Additionally, the classical molecular dynamics (MD) simulations are applied to gain insights into the Na-ion transport properties in both NMFO and B-NMFO structures, which suggest that the B-doped NMFO exhibits enhanced Na-ion diffusivity compared to the pristine NMFO sample. 

\section{\noindent ~Experimental and theoretical}

{\noindent ~ {\it Materials synthesis}:} 

The P2-type layered NMFO and B-NMFO samples are prepared using high purity precursors ($\geq$99.99\%) from Sigma without any purification process through a typical sol-gel method using citric acid as a chelating agent \cite{XuCEC14}. The precursor materials such as NaNO$_{3}$, Mn(CH$_{3}$COO)$_{2}$.4H$_{2}$O, Fe(NO$_{3}$)$_{3}$.9H$_{2}$O, H$_{3}$BO$_{3}$ and C$_{6}$H$_{8}$O$_{7}$ in a stoichiometric ratio (0.66:0.8:0.1:0.1:1) with 5\% Na excess were dissolved in a 40 ml distilled water and stirred for 1 hr at 25$^\circ$C until to get a clear solution. Further, a 20 ml of citric acid solution was made in order to form a gel matrix with the above solution. Then, the mixed solution was kept at 80$^\circ$C (oil bath) under a constant stirring for 5 hr leading to the reddish gel which was subjected for overnight aging. After that, as formed gel was dried in oven at 120$^\circ$C for 24h and then ground and made into discs, which were subjected to a two stage heating processes (grinding at each steps) at 400 and 900$^\circ$C for 5 and 15 hr, respectively to obtain P2 phase and finally kept the materials in the glove box having inert environment.  

{\noindent ~  {\it Physical characterization details}:} 

To investigate the crystallographic structure of the active material, the room temperature X-ray diffraction measurements (Panalytical X'pert$^3$) are performed with CuK$\alpha$ radiation (1.5406\AA) at scan rate 2$\degree$/minute in 2$\theta$ range of 10 to 60$\degree$ and high resolution transmission electron microscopy (HR-TEM) using a Tecnai G2 20 system. The energy dispersive x-ray spectroscopy (EDX) measurement is performed by using RONTEC’s EDX system Model QuanTax 200 to find the elemental composition and particle distribution across the sample. The Raman spectroscopy data are collected to understand the microstructure of the active material using a Renishaw inVia confocal Raman microscope 2400 lines/mm grating at a wavelength of 532 nm with a laser power of 1 mW on the sample. The X-ray photo-electron spectroscopy (XPS) measurements are carried out with AXIS Supra system having monochromatic Al K$\alpha$ source (1486.6 eV) and energy resolution of $\approx$0.5 eV, to probe the electronic states of individual elements. The  binding energy (BE) of C 1$s$ core-level spectrum at 284.6~eV is used to calibrate individual peak positions. We use Voigt function and Tougaard method for deconvolution of the core-level peaks and background subtraction, respectively.

{\noindent ~  {\it Coin cell fabrication details}:} 

In order to prepare the electrodes, we first make slurry by taking the active material, carbon black as the conducting source and polyvinylidene fluoride (PVDF) as the binder in the ratio of 80 : 10 : 10 (by weight) and mixed in N-methyl-2-pyrrolidone (NMP) solvent. The as-prepared slurry is coated on an aluminium foil using a doctor blade method with active material mass loading of 2--3 mg cm$^{-2}$ and then vacuum heated at 120\degree C for 12 hrs. After heating, the electrodes are cut into 12 mm diameter discs and again dried under vacuum at 80\degree C before inserting in glove box. We use 1M NaPF$_6$ electrolyte dissolved in EC and DEC (50 : 50 by volume), sodium metal foil as a reference electrode, a glass fibre (GB100R) as separator and 12 mm diameter of the active electrode to fabricate CR2032 type coin cells inside an argon gas-filled glove box (UniLab Pro SP from MBraun) having O$_2$ and H$_2$O level less than 0.1~ppm. 

{\noindent ~  {\it Electrochemical characterization details}:} 

The Biologic VMP-3 system is used to conduct the CV at multiple scan rates, GCD at different current rates, EIS and GITT measurements. The EIS measurements are carried out within a frequency range of 10~mHz to 100~kHz, with a maximum ac voltage of 10~mV at the open circuit voltage of the cells. A Neware battery cycler (BTS400) is used to collect the GCD profiles in the voltage window of 1.5--4.2~V (vs Na$^{+}$/Na) for rate capability test and long cycling. For the GITT measurements, a constant current rate of 0.1~C was provided with a current pulse of 10~min followed by an open circuit relaxation of 1~hr. An open source tool based on MATLAB was used to obtain the DRT (distribution of relaxation time) results of the related EIS spectrum at different temperatures. Further, the softBV-GUI-v131 software employs bond-valence-based empirical force fields to calculate the activation barrier and conduction pathways of Na-ions. 

{\noindent ~  {\it Computational Methodology}:}  

The  bulk structure of Na$_{0.66}$Mn$_{0.8}$Fe$_{0.2}$O$_{2}$ (NMFO) is modeled based on the stoichiometry of the sample. Since atomistic simulations require whole numbers of atoms, the composition is approximated by removing Na atoms and substituting Mn atoms with Fe in the original NaMnO$_{2}$ crystal. The primitive unit cell of bulk NaMnO$_{2}$, which has a hexagonal crystal system with the P6$_3$/$mmc$ space group, is taken from the Materials Project database (material ID: mp--971647) \cite{JainAPL13}. This approach preserves the crystal system as per the experiments, and the lattice parameters are further adjusted based on our XRD data. The final NMFO structure is then optimized using high-precision density functional theory (DFT) calculations. The first principles calculations are performed using the Vienna Ab initio Simulation Package (VASP version 6.2) \cite{KressePRB96, KressePRB93}. The lane wave DFT within the projector-augmented wave (PAW) \cite{BlochlPRB94} framework method is utilized to describe electron-ion interactions. A plane-wave cutoff energy of 520~eV is used. The exchange-correlation functional is treated using the Perdew-Burke-Ernzerhof (PBE) generalized gradient approximation (GGA) \cite{PerdewPRL96}. An on-site Coulomb interaction (U) of 4.0 eV is applied to the 3$d$ electrons of the transition metals Fe and Mn using the Hubbard U correction model  \cite{DudarevPRB98}. The U values are estimated using the linear response approach, a standard method for estimating U values in the range of 4.0--5.0 for Fe and Mn in such similar layered transition metal oxides and spinel oxides~\cite{JainPRB11, ZhouPRB04}. The spin-polarized calculations are performed to fully capture the system’s magnetic and electronic properties. The electronic minimization used the blocked Davidson algorithm \cite{DudarevPRB98}, with a total energy and force convergence criterion of 10$^{-6}$~eV and -0.01~eV/\AA, respectively.  The ionic relaxation is performed using the conjugate gradient algorithm, optimizing both atomic positions and cell volume. The Gaussian smearing with a width of 0.05~eV is applied for stable convergence. These setup provide an accurate and efficient approach, as suggested by our previous studies \cite{SethAEM25, SinghCEJ23} for studying the electronic and structural properties of the similar layered oxide materials. In order to measure the amount of charge transfer for individual atoms and provide a more direct indication of the impact of boron atom on the charge distribution, the change of Bader charges following doping with boron atom is determined, as described by Henkelman et al. \cite{HenkelmanCMS06}. 

\begin{figure*}
\includegraphics[width=7.3in]{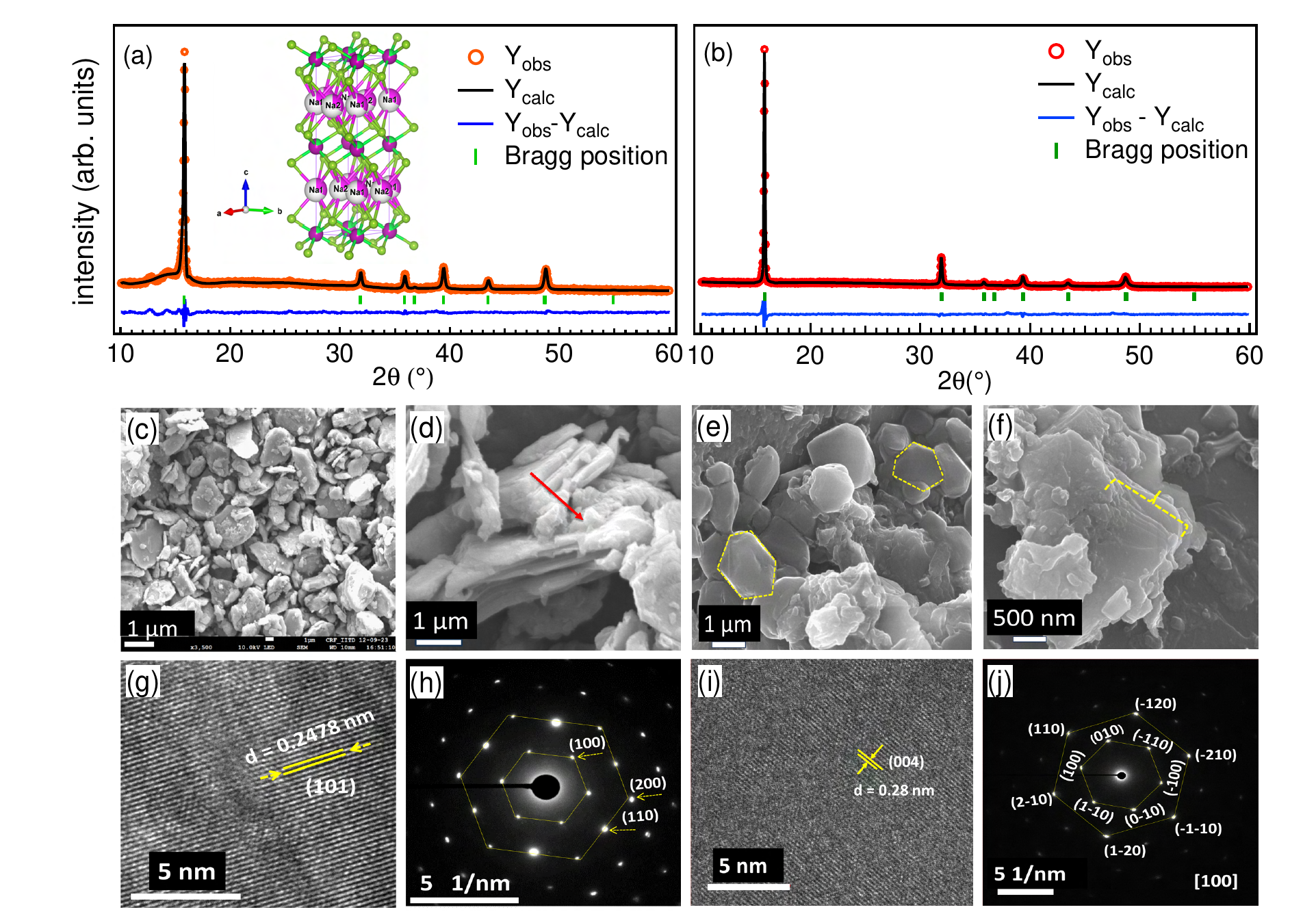}
\caption {The Rietveld refinements (black lines) of X-ray diffraction patterns (red color) of the (a) NMFO and (b) B-NMFO samples with corresponding crystal structures in the inset of (a). The FE-SEM images showing submicron particles of (c) NMFO and (e) B-NMFO samples with the magnified view showing flake like layered patterns in (d, e), respectively. The HR-TEM images showing 2D lattice fringes with inter-planar distance in (g, i) and the SAED patterns with respective indexed planes in (h, j) for the NMFO and B-NMFO samples, respectively.}
 \label{Fig-1}
\end{figure*}

Moreover, the molecular dynamics (MD) simulations are performed to determine the Na-ion transport in both NMFO and B-NMFO supercell structures. The interatomic Buckingham \cite{BuckinghamPSC97} pair-potential model is applied using the code available in LAMMPS (ver. June 2019) \cite{Plimpton95} software. The full potential equation and parameters are provided in Table~S1 of \cite{SI}. A supercell of size 5$\times$5$\times$5 is modelled using the DFT optimized primitive unit cell, as discussed above, to study ion-transport in bulk NMFO structure. The energy minimization is performed to relax the structure, followed by the equilibration for 1~ns using canonical ensemble (NVT) at 300 K. The final production runs are performed using an isothermal-isobaric (NPT) ensemble for 10 ns duration with time-step of 1~fs. To control the temperature fluctuations a Nosé-Hoover thermostat with a damping factor of 1000~fs is used. An average 3-dimensional transport coefficients of Na-ions are determined from the slope of mean square displacement (MSD) versus time plot, following Einstein’s relations \cite{AllnattJPC82}:
\begin{equation}
D_{\text{Na}^+} = \frac{1}{6t} \langle MSD \rangle
\end{equation}                                                   
where, $\langle {\it MSD} \rangle$ is $\langle |\mathbf{r}(t) - \mathbf{r}(0)|^{2} \rangle$; {\it t} is time; and  {\it D$_{Na^{+}}$} is self-diffusion coefficient of Na-ions. 

\section{\noindent ~ Results and discussion:}

The Rietveld refined XRD profiles of both the samples with the Bragg's positions and the respective crystal structure are shown in Figs.~\ref{Fig-1}(a, b). The refinement depicts the hexagonal layered crystal structure with a space group of P6$_3$/$mmc$ \cite{XuCEC14}. All the diffraction peaks are well defined and matched [JCPDS. No. 194]. Further, the refinement output confirms the lattice parameters $a=b=$ 2.8880 \AA, $c=$ 11.2285 \AA, $\alpha$ = 90\degree , $\beta$ = 90\degree and $\gamma$ = 120\degree and V = 81.1048 \AA$^{3}$ with a goodness of fit value $\chi$$^{2}$ = 1.37 for the NMFO sample. The detailed refinement parameters are provided in Table~S1, which shows Mn and Fe atoms take the octahedral 2$a$ site and O atoms occupy the 4$f$ site. Meanwhile, there are two different Na$^{+}$ sites in the local prismatic structure where the one is located at edge-shared Na$_{e}$ (2$b$) and the other is occupied at face shared Na$_{f}$ sites (2$d$) between higher and lower MnO$_{6}$ octahedral layers, respectively \cite{FengSmall23}. The extracted lattice parameters of the B-NMFO sample are found to be $a=b=$ 2.8951 \AA, $c=$ 11.21146 \AA, and V = 81.3863 \AA$^{3}$, which are essentially similar to the NMFO sample. Here, the B$^{3+}$ mainly occupies the tetrahedral interstice or trigonal sites of packed oxygen in the transition metal (TM) layer and sodium layer to form (BO$_{4}$)$^{5-}$ and (BO$_{3}$)$^{3-}$ polyanionic structures, when boron is doped in the bulk structure \cite{LiAFM14, PanJPS16}. Owing to the small ionic radius of B$^{3+}$ (0.27\AA) than the metal ions, there are no significant changes in the cell parameters after boron doping. In Fig.~\ref{Fig-1}(c), the FE-SEM image of as-prepared samples show the sub-micrometric irregular slabs with an average particle size of 2--3 $\mu${\it m} and the red arrow in Fig.~\ref{Fig-1}(d) depicts the flake like particles with clearly visible layered pattern present in the samples \cite{YangAFM20}. The layered structure provides large surface area for the electrode-electrolyte interaction, which helps in the ease migration of Na-ions. Intriguingly, the B-NMFO sample exhibits quasi-polygonal shape particles with average size of 0.7-1.8~$\mu${\it m} , as shown in Fig.~\ref{Fig-1}(e). These quasi-polygonal (asymmetrical and moderately hexagonal) shape provide better electrochemical kinetics in achieving higher capacity \cite{MathiyalaganIJER22}. The primary reason of obtaining moderately hexagonal and small sized particles may involve high temperature (900\degree C) calcination and boron doping \cite{MathiyalaganIJER22}. The boron doping may reduce the surface energy of some crystal planes by inhibiting the growth in that particular direction and results in the modification of primary particles \cite{MathiyalaganIJER22}. Fig.~\ref{Fig-1}(f) shows the slipovers with layered pattern for the B-NMFO sample. Also, the HR-TEM images in Figs.~\ref{Fig-1}(g, i) clearly depict the 2D fringes with inter-planar distance of 0.24  and 0.28~nm of the corresponding (101) and (004) crystal planes of both the samples, respectively. Moreover, the SAED patterns in Figs.~\ref{Fig-1}(h, j) reflect the hexagonal symmetry for the layered samples and the respective atomic planes are indexed, confirming the P2 hexagonal phase. This is well corroborated by the lattice fringes, see Figs.~\ref{Fig-1}(g, i), and attribute well with crystalline structure of both the samples \cite{FengSmall23, ManikandanEA16}. 

\begin{figure}[h]
\includegraphics[width=3.55 in]{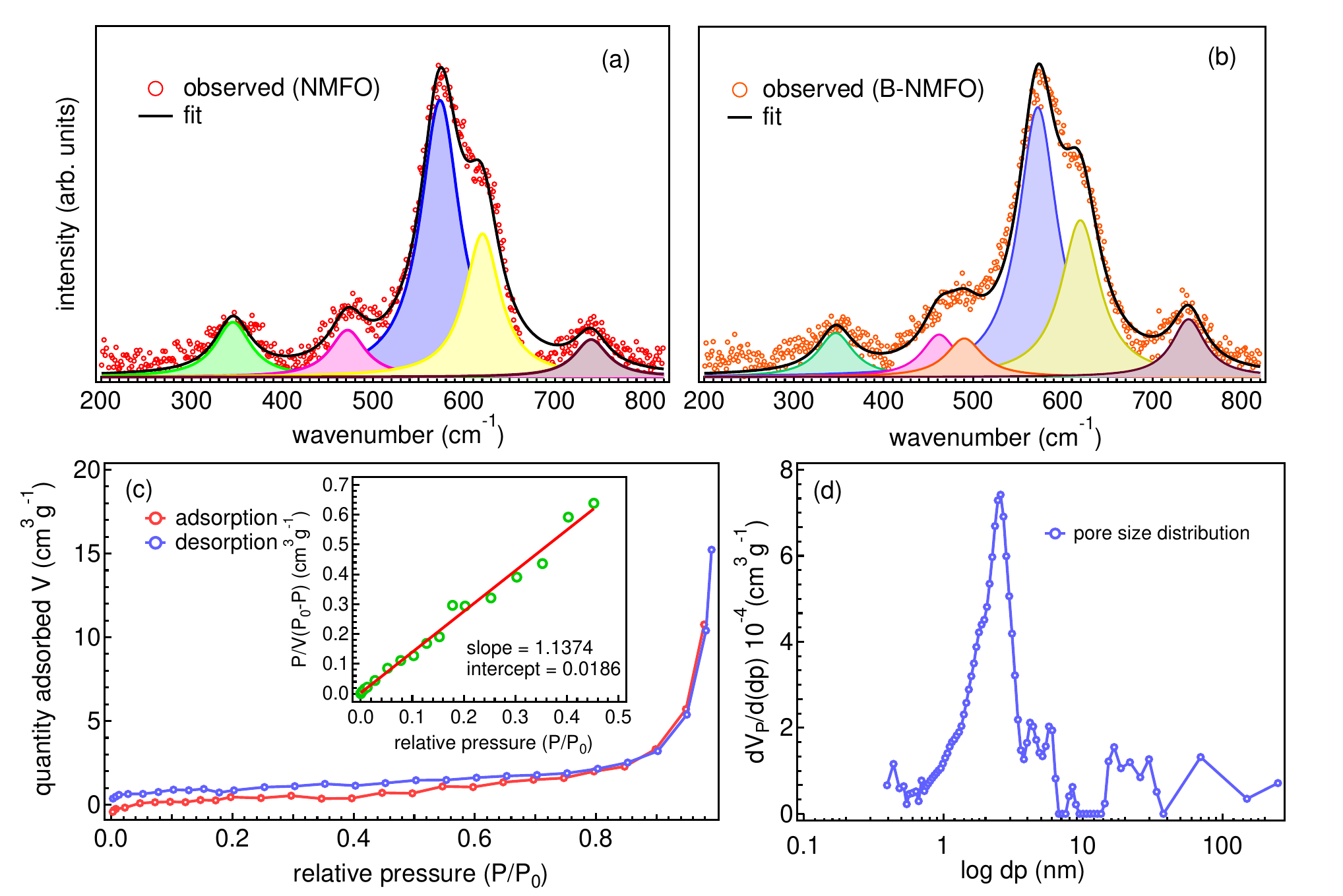}
\caption {The Raman spectra in a range of 200--820 cm$^{-1}$ for the (a) NMFO and (b) B-NMFO samples. (c) The BET adsorption-desorption isotherms and (d) the pore size distribution of the NMFO sample.}
 \label{Fig-2}
\end{figure} 

Further, the Raman spectra are recorded in a range of 200--800 cm$^{-1}$ to investigate the co-ordination geometry, Jahn-Teller distortion, and short range structural disorder for both samples. Herein, the pristine sample contains both Mn$^{3+}$ and Mn$^{4+}$ cations, which results variation in MnO$_{6}$ octahedral geometry due to Jahn-Teller distortion arising from high spin Mn$^{3+}$ ions (3d$^{4}$). According to the group theory formulations, P2--hexagonal symmetry is predicted to show five Raman active modes, A$_{1g}$ + 3E$_{2g}$ + E$_{1g}$ \cite{VassilevaAST10} where three of them are from Mn--O vibrations. In Figs.~\ref{Fig-2}(a, b), the A$_{1g}$ peak positioned at higher wavenumber, 618 cm$^{-1}$ is assigned to symmetric stretching vibration $\nu$$_{2}$ (Mn--O) whereas the high intense E$_{2g}$ band appeared at 573 cm$^{-1}$ is attributed to the stretching vibration of $\nu$$_{3}$ (Mn--O) in basal planes of Mn-O$_{6}$ octahedron \cite{YangAFM21}. The dominancy of these features can clearly be related to the formation of P2 hexagonal family. Further, the O--Mn--O bending vibrations are detected at 345 and 473 cm$^{-1}$ which can introduce Na site ordering as the Mn--O layers are well separated by Na--O bonded layers. However, a weak peak at 737 cm$^{-1}$  is also originated due to the  stretching vibration Mn--O \cite{ZhangJCIS24, BernardiniJRS19}. For the B-NMFO sample, a new weak feature originates at 462 cm$^{-1}$ representing the vibrational mode of B--O bond \cite{AngelCM19}. Further, Fig.~\ref{Fig-2}(c) illustrates the N2 absorption-desorption isotherms of the NMFO sample, which shows a hysteresis (within a P/P0 range of 0.2 to 0.8) depicting the type IV behavior (H3 type) due to the characteristic nature of mesoporous structures.  In general, the H3 loops infer aggregates of plate-like particles or slit-shaped pores and typically do not show a clear limiting adsorption plateau at high relative pressure \cite{ThommesPAC15}. The sharp uptake near high relative pressure suggests capillary condensation in mesopores, while the low-pressure region shows limited micropore filling and multilayer adsorption \cite{FuEA19}. The specific surface area of NMFO determined via Brunauer-Emmett-Teller (BET) method is found to be 3.7 m$^{2}$g$^{-1}$ by using the linear relationship between P/V(P$_{0}$--P) and P/P$_{0}$, as shown in the inset of Fig.~\ref{Fig-2}(c). This along with the porous features facilitates the infiltration of electrolyte and prevents the agglomeration in the cathode \cite{ShenJMCA18}. Further, the pore size distributions analysis by Barrett-Joyner-Halenda (BJH) method is presented in Fig.~\ref{Fig-2}(d). Here, we observe a majority of pore size as  2.5~nm with the average pore diameter and pore volume as 20.13~nm and 2.24$\times$10$^{-2}$ cm$^{3}$ g$^{-1}$, respectively. Moreover, the energy dispersive X-ray spectrum reveals the existence of Na, Mn, Fe and O elements in their expected respective ratios, as shown in Fig.~S1 of \cite{SI} where the elemental distribution shows a uniform segregation of Na, Mn, Fe and O on the layered structure.  
 
\begin{figure*}
\includegraphics[width=7in]{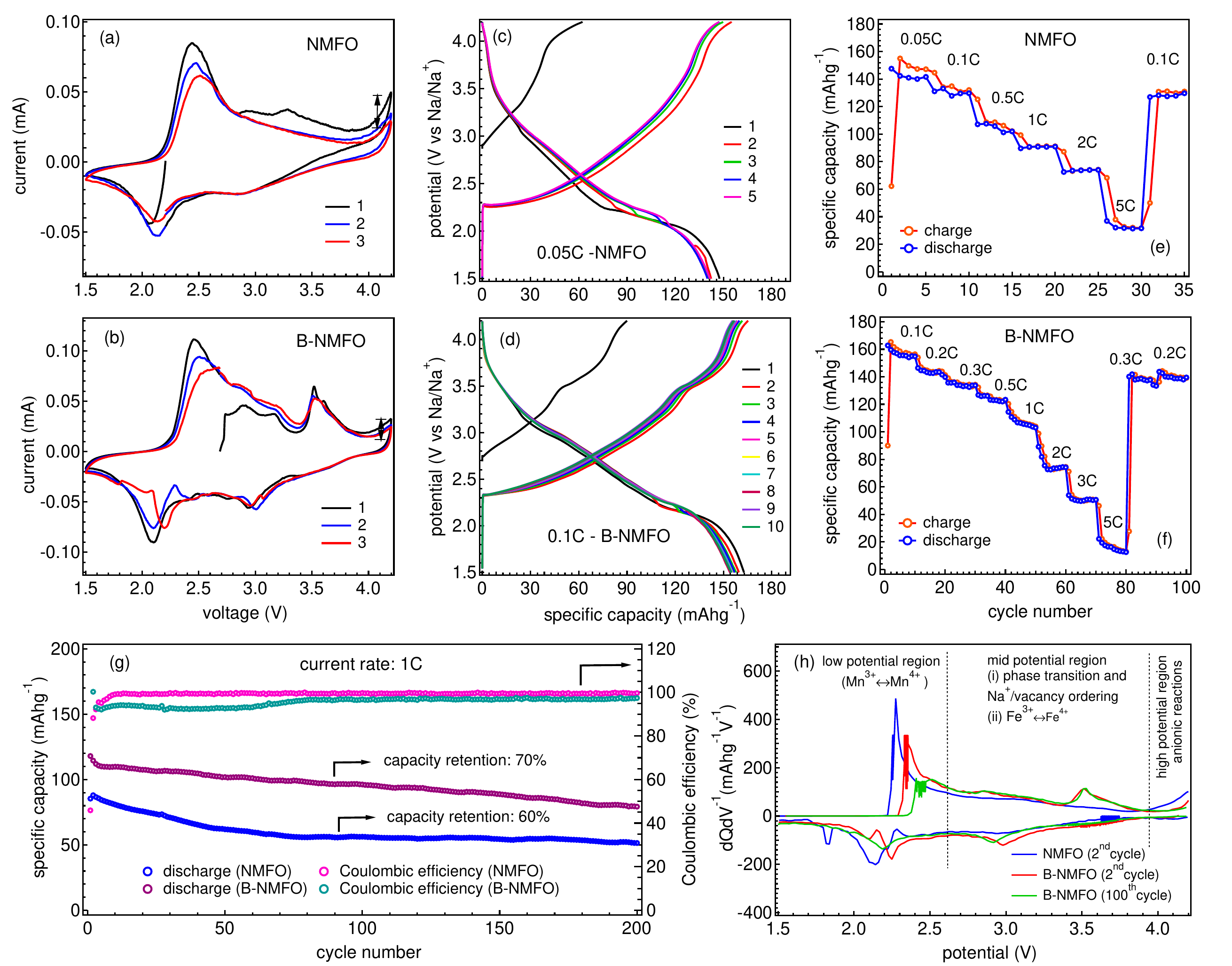}
\caption {The electrochemical measurements of NMFO and B-NMFO cathodes: (a, b) the cyclic voltammetry curves at a scan rate of 0.05 mVs$^{-1}$ for 3 cycles in a voltage window of 1.5--4.2~V,  (c, d) the galvanostatic discharge-charge (GCD) voltage profiles at a current rate of (c) 0.05~C for 5 cycles for NMFO, and (d) 0.1~C for 10 cycles for B-NMFO. (e, f) The rate capability measurement at different current rates in a voltage window of 1.5--4.2~V for the (e) NMFO up to 5 cycles each, and (f) B-NMFO up to 10 cycles each. (g) the long cycling measurement of NMFO and B-NMFO at current rate of 1~C up to 200 cycles, (h) the dQ/dV versus voltage curves for 2$^{nd}$ and 100$^{th}$ cycles of B-NMFO and 2$^{nd}$ cycle of the NMFO.}
 \label{Fig-3}
\end{figure*}

\begin{figure*}
\includegraphics[width=1\textwidth]{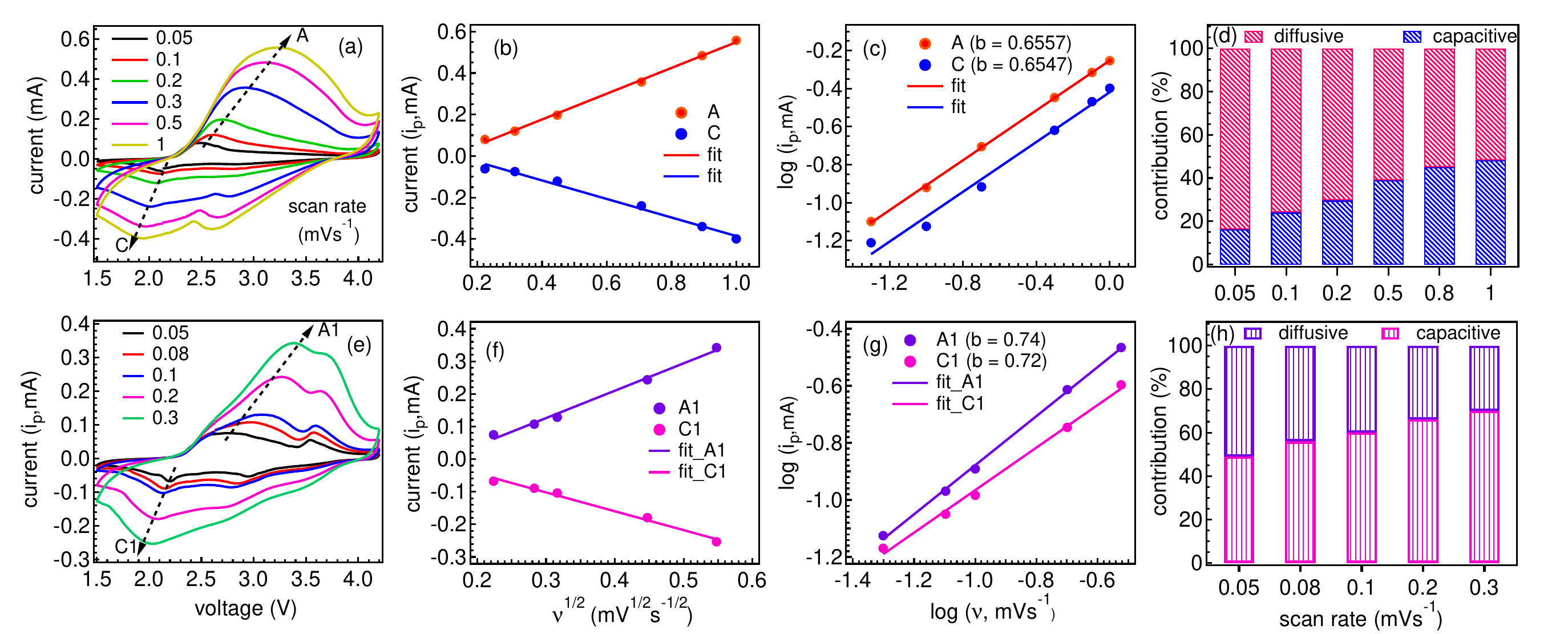} 
\caption {The Cyclic voltammetry curves at different scan rates in a voltage window of 1.5--4.2~V, showing anodic (A, A1) and cathodic (C, C1) peaks for the (a) NMFO and (e) B-NMFO cathodes. The linear relation between peak current  $i_{p}$ and square root of scan rate for different scan rates for the (b) NMFO (f) B-NMFO. The linear fitting between log  $i$ vs log $\nu$ for the (c) NMFO (g) B-NMFO. The capacitive and diffusive contribution at different scan rates for the (d) NMFO (h) B-NMFO.}
 \label{Fig-4}
 \end{figure*}

In order to understand the redox mechanism during Na-ion de-/insertion in the host structure, we carried out the cyclic voltammetry (CV) measurements in a voltage window of 1.5--4.2~V at scan rate of 0.05 mVs$^{-1}$, as shown in Figs.~\ref{Fig-3}(a, b) for three cycles. We can clearly observe low potential redox pairs below 2.5~V depicting the Mn$^{3+/4+}$ redox reaction \cite{NayakAEM22, ZuoNAT21}. The weak intensity of the cathodic peak suggests partial irreversibility of the Mn$^{4+}$ reduction phenomenon because some Mn ions can retain well in the 4+ oxidation state. Notably, the minimal potential deviation (voltage polarization) underscores the reversible Na-ion kinetics. There are small redox peaks below 3~V, which correspond to the structural rearrangement of the layers during de-/sodiation \cite{HemalathaDT18}. In Fig.~\ref{Fig-3}(b), the weak redox events in the range of 2.5--4 V comprise of complex phase transitions incurred by the Na$^{+}$ vacancy ordering and Fe$^{3+/4+}$ reaction \cite{BoivinAEM22}. Intriguingly, a distinct and reversible redox pair at 3.5/3~V provides an additional capacity in the doped cathode \cite{SuAEM24}. The small oxidation peak at $\sim$ 4.1~V due to oxygen redox reaction (O$^{2-}$/O$^{n-}$) signifies the electrochemical stability of used electrolyte (1M NaPF$_{6}$ in carbonate solvents) within the operational voltage window \cite{EshetuAEM20}. Further, the Na$^{+}$ storage behavior is investigated by measuring the Galvanostatic charge-discharge curves, as represented in Figs.~\ref{Fig-3}(c, d). A couple of plateaus at 2 and 4~V are visible in the discharge profiles, which are in well accordance with the CV results. The excellent initial discharge capacity of 148~mA hg$^{-1}$ is observed at 0.05~C for the pristine cathode [see Fig.~\ref{Fig-3}(c)], whereas the B-NMFO cathode provides much improved initial discharge capacity of 163~mA hg$^{-1}$ at 0.1~C [see Fig.~\ref{Fig-3}(d)], which may be due to the secondary capacity contribution from the 3.5~V plateau 
corresponds to the P2 to OP4 transition providing additional capacity in B-NMFO cathode \cite{SuAEM24, LuoAMI22, SiriwardenaCEC21}. However, the large hysteresis difference in redox potentials of first cycle and the subsequent cycles is basically arises due to the cathode electrolyte interface (CEI) and the structural ordering during oxygen redox reactions. Also, the 1$^{st}$ cycle charging capacity is much lower for both the cathodes, which may be due to the fact that the un-cycled material have 0.66 Na$^{+}$ ion per formula unit, as a result the Mn$^{3+}$/$^{4+}$ redox is not fully involved in the reaction. While during the first discharge process, maximum amount of  Mn$^{4+}$ ions are reduced to the trivalent state and the material gets enough Na$^{+}$ ions. Thus, in the second charging process, the Mn$^{3+}$/$^{4+}$ redox is fully involved and resulted in much improved and stable capacity in subsequent cycles. 

Notably, in the rate capability tests shown in Figs.~\ref{Fig-3}(e, f), the B-NMFO cathode exhibits better performance as compared to the NMFO particularly at lower current rates upto 2 C. Notably, the excellent rate capability, i.e., better than 95\% retention of the initial capacity is achieved for both the cathodes after switching the cell from 5~C to 0.1--0.2 C rates, see Figs.~\ref{Fig-3}(e, f). To some extend better performance of B-NMFO cathode at below 2 C rates may originates from the high bond energy of B--O in comparison to Mn--O and Fe--O, which prevents the irreversible loss of lattice oxygen from the host structure and also the B-doping can tune the electronic structure by lowering O-2$p$ band with a further enhancement of oxygen stability in layered compounds \cite{LiCM17,DangJEC24}. Subsequently, we tested cyclic stability at 1~C, as shown in the Fig.~\ref{Fig-3}(g), which show the retention of around 60\% and 70\% for the NMFO and B-NMFO, respectively, after 200 cycles, maintaining Coulombic efficiency 98-100\%. The dQ/dV versus voltage curve at 2$^{nd}$ cycle for both the cathodes and at 100$^{th}$ cycle for the B-NMFO cathode (after rate capability test) are presented in Fig.~\ref{Fig-3}(h). Here, for the 2$^{nd}$ cycle data, the low potential peaks attribute to Mn$^{3+}$/$^{4+}$ redox activity and the potential gap between these peaks is higher (0.134 V) for NMFO as compared to the values (0.096 V) for the B-NMFO suggesting its better reversibility. In case of 100$^{th}$ cycle, the degradation of Mn peak is observed due to the Mn-dissolution in subsequent cycles. The peaks in the mid potential region suggest two possibilities, which include phase transition from P2 to OP4 and redox reaction of Fe$^{3+/4+}$. Additionally, the high potential range suggest small amount of anionic redox reaction due to structural change and O-O dimerization \cite{BoivinAEM22, SuAEM24}.  

Now we move to the investigation of diffusion kinetics through cyclic voltammetry (CV) and the galvanostatic intermittent titration technique (GITT) analysis. Figs.~\ref{Fig-4}(a, e) show the CV curves for the NMFO and B-NMFO cathodes at different scan rates in the voltage range of 1.5--4.2~V (vs. Na/Na$^{+}$), highlighting cathodic and anodic peaks, which gradually shift to lower potential and higher potentials, respectively, as the scan rate increases from 0.05 to 1 mV s$^{-1}$. For further analysis, in Figs.~\ref{Fig-4}(b, f), we plot the peak current i$_{p}$ versus square root of scan rate ($\nu$$^\frac{1}{2}$) for the anodic and cathodic couples, marked as A, C for the NMFO cathode, and A1, C1 for the B-NMFO cathode, respectively. To calculate the Na- ion diffusion coefficient values, the relationship between $i_{p}$ versus $\nu$$^\frac{1}{2}$ is described by the Randles-Sevcik equation for the diffusion controlled electrochemical reaction, as written below \cite{PatiJMCA22, SinghCEJ23}:
 \begin{eqnarray}
 i_{p} = (2.69 \times 10^{5}) A D^{\frac{1}{2}}  C n^{\frac{3}{2}} \nu^{\frac{1}{2}},
 \end{eqnarray}
here, $i_{p}$ is the peak current (mA), A is the area of the electrode (cm$^{2}$), C is the bulk concentration of the Na ions in the electrode (mol cm$^{-3}$), $n$ is the no of the electrons transferred in an electrochemical reaction, $\nu$ is the scan rate (mV s$^{-1}$) and $D$ is the diffusion coefficient of the sodium ions in the cathode (cm$^{2}$ s$^{-1}$). The slope of a linear fit to the plots of $i_{p}$ versus $\nu$$^\frac{1}{2}$, see Figs.~\ref{Fig-4}(b, f), is used to calculate the $D$ values for both cathodic and anodic peaks. The calculated values of $D$ for cathodic and anodic peaks found to vary in the range of 10$^{-8}$ to 10$^{-10}$ cm$^{2}$s$^{-1}$, which are consistent and considered for relatively faster sodium-ion diffusion kinetics in cathode materials \cite{YouAEM17}.  

Moreover, the energy storage mechanism of the Faradaic storage device exhibits two distinct processes, including diffusion-controlled (battery) and surface-controlled (capacitive). To evaluate their contributions, a linear relationship between the logarithimic of peak currents and different scan rates is presented in Figs.~\ref{Fig-4}(c, g) for both NMFO and B-NMFO cathodes, respectively, which can be described by the following equation: 
\begin{eqnarray}
i = a\nu^{b}
\end{eqnarray}
where {\it a} and {\it b} are adjustable parameters. As per the universal notion, the {\it b} value around 0.5 signifies that the dominant process is diffusion-controlled (battery), whereas the {\it b} value of nearly 1 can be attributed to surface-controlled (capacitive) process. The {\it b} parameter represents the slope by linear fitting the data points in Figs.~\ref{Fig-4}(c, g). The {\it b} values for both anodic (Peak A) and cathodic peaks (Peak C) are 0.65 for NMFO cathode, and in case of the B-NMFO cathode, the {\it b} values of anodic and cathodic peaks are found to be 0.74 and 0.72, respectively. These values indicate that the charge storage mechanism is combination of diffusive and capacitive, i.e., of pseudocapacitive nature. The slightly higher {\it b} values in B-NMFO suggests more number of surface contributed reactions than the NMFO, which attributes additional capacity in the B-NMFO cathode. Furthermore, the combination of diffusion-influenced reaction kinetics and surface-originating capacitive distribution at a specific voltage can be quantitatively calculated using the following equation \cite{SapraAMI24, PatiJPS24}:
\begin{eqnarray}
i(V)/\nu^{1/2} = k_{1} \nu^{1/2}+k_{2}
 \end{eqnarray}
 where, {\it k}$_{1}$$\nu$ and {\it k}$_{2}$$\nu$$^{1/2}$ denote the surface capacitive process and diffusion-limited redox reaction at a specific scan rate, respectively. For instance, at a sweep rate of 0.2 mVs$^{-1}$, the shaded area, see Fig.~S2 of \cite{SI}, corresponds to the surface capacitive contribution in the total capacity. In order to demonstrate the impact of capacitive/diffusive growth at successive scan rates, histograms are plotted in Figs.~\ref{Fig-4}(d, h). With an increase in the scan rate, the capacitive ratio increases in both the cathodes, indicating the dominant role of the surface-capacitive effect at high scan rates.

 \begin{figure}[h]
\includegraphics[width=0.45\textwidth]{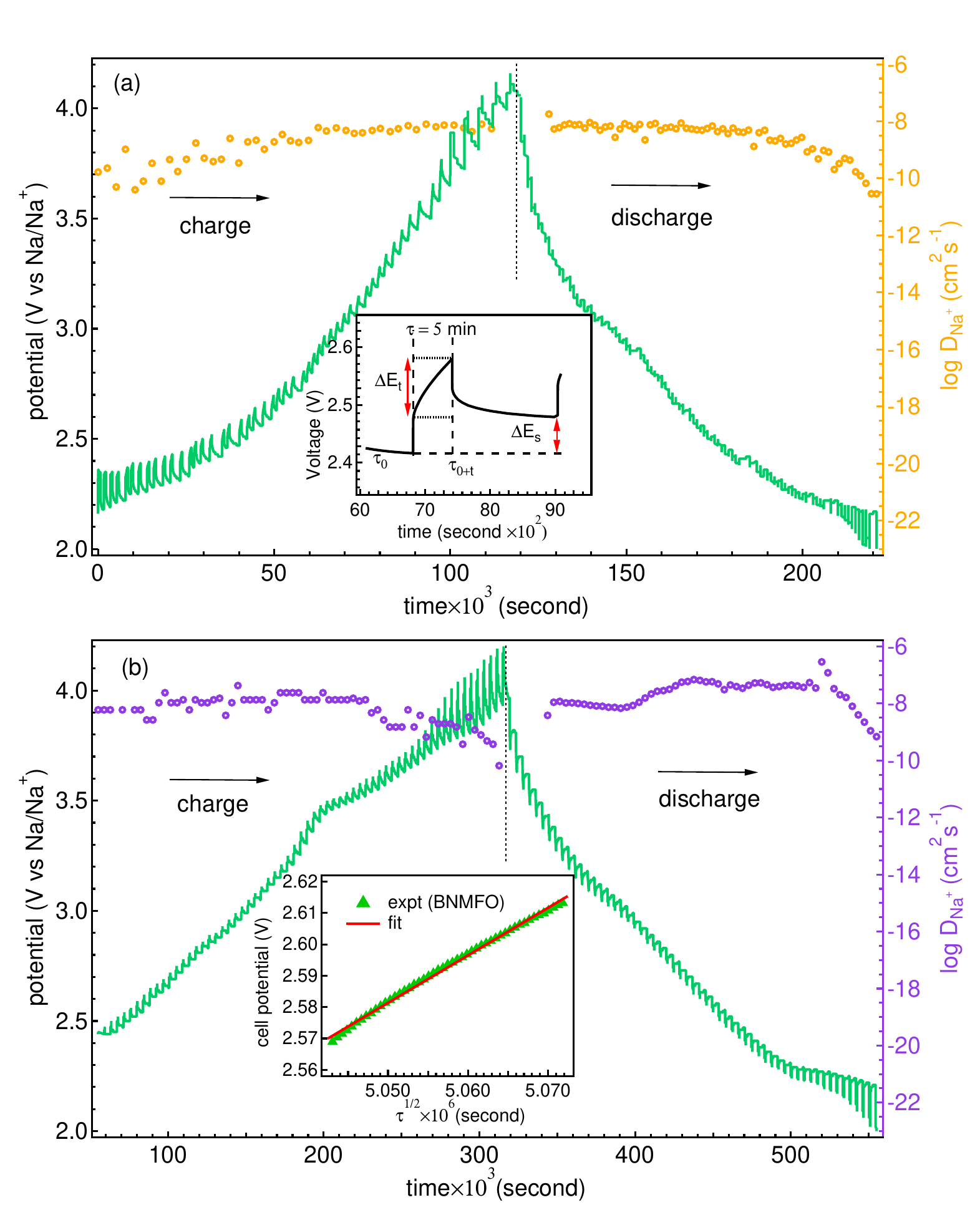}
\caption {The GITT measurements at 0.1~C in a voltage window of 2.0--4.2 V during 2$^{nd}$ cycle with logarithimic diffusion-coefficient profile of (a) the NMFO cathode with the inset showing a schematic labeling of different parameters of a single titration curve before, during and after application of a current pulse for 10 min, (b) the B-NMFO cathode with inset presenting linear relation between cell potential and $\tau$$^{1/2}$ .}
 \label{Fig-5}
 \end{figure}

 \begin{figure*}
\includegraphics[width=1\textwidth,height=13cm]{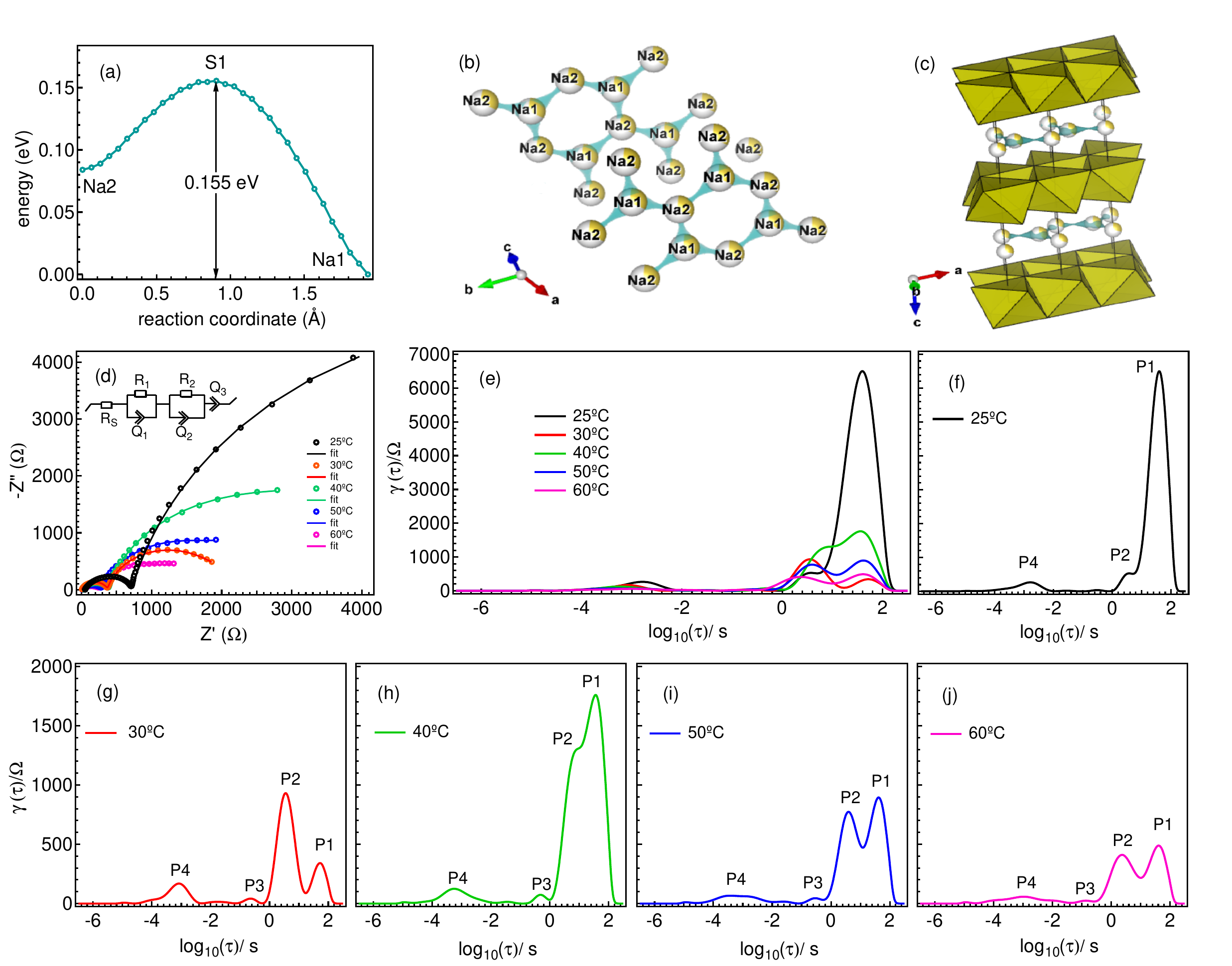}
\caption {The Na ion migration and the DRT mechanism of B-NMFO cathode: (a) the energy and reaction co-ordinate plot showing the energy barrier between the two Na-sites, (b) the 2D conduction pathways of Na1 and Na2 in the $ab-$plane, (c) the 3D view of Na-ion conduction paths shown by blue colored isosurfaces in between the gallery of yellow color octahedral slabs, (d) the EIS recorded in a temperature range of 25--60\degree C, the DRT profiles of EIS spectra recorded at (e) different temperatures, (f) 25\degree C, (g) 30\degree C, (h) 40\degree C, (i) 50\degree C, (j) 60\degree C.}
 \label{Fig-6}
 \end{figure*}

Furthermore, the diffusion coefficient at specific voltages can be obtained through GITT analysis by using Fick's second law of diffusion \cite{DangJEC24}:
\begin{equation}
D_{Na^{+}}= \frac{4}{\pi \tau}\left[\frac{m_B V_M}{M_B A}\right]^2\left[\frac{\Delta E_s}{\Delta E_t}\right]^2;\tau=L^{2}/D_{Na^{+}}
\end{equation}
where $\tau$ is the constant current pulse time, {\it A} represents the contact area between the electrode and electrolyte (here the geometrical surface area of the electrode, 1.13 cm$^{2}$), while {\it m}$_{B}$ , {\it V}$_{M}$, and {\it M}$_{B}$ are the actual mass (g), the molar volume (cm$^{3}$ mol$^{-1}$), and the molar mass (g mol$^{-1}$) of the electrode materials, respectively. The $\Delta${\it E}$_{s}$ is the voltage difference from the steady-state voltage during a single-step GITT process and $\Delta${\it E}$_{t}$ attributes the change in potential during the application of current pulse. The GITT profiles of the NMFO and B-NMFO cathodes during the entire charging and discharging process are shown in Figs.~\ref{Fig-5}(a, b), and a single-step titration curve with all the parameter details is presented in the inset of Fig.~\ref{Fig-5}(i). Here, in both the cases a 10 min ($\tau$ = 10 min) of current pulse is applied to the system followed by a OCV stand (rest time) of 1 hr during charging-discharging at a constant current rate of 0.1~C. The GITT profiles are consistent with the GCD profiles in Figs.~\ref{Fig-3}(c, d), confirming the viability of the data for further analysis. The extracted values of diffusion, using equation~5, are plotted as log of D$_{Na+}$, as a function of cell potential, on the right scale of Figs.~\ref{Fig-5}(a, b), which show a comparable and stable behavior in the potential range of 2-4.5 V and consistent with the CV analysis presented above. 

\begin{figure*}
\centering
\includegraphics[width=1\textwidth]{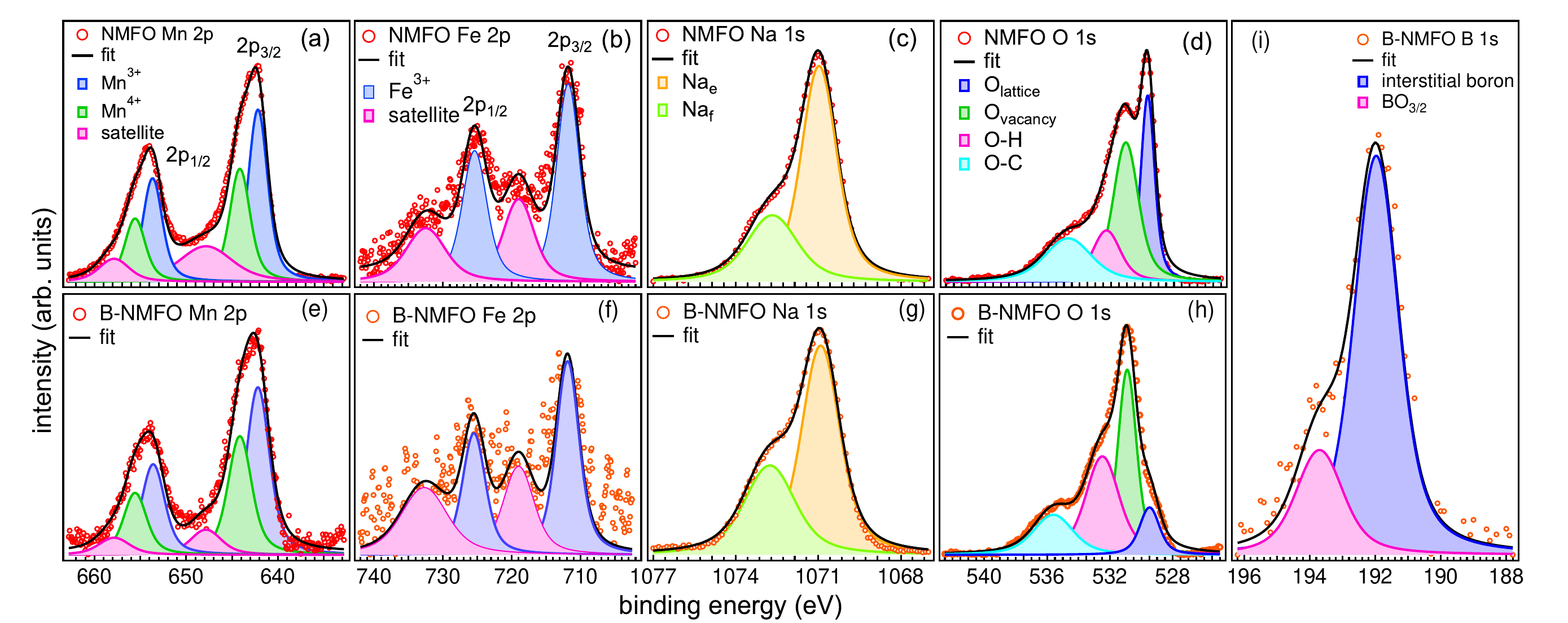}
\caption {The core level XPS spectra of Mn 2$p$, Fe 2$p$, Na 1$s$ and O 1$s$ of (a--d) NMFO, (e--h) B-NMFO samples, (i) the B 1$s$ of B-NMFO sample along with the deconvoluted components.}
\label{Fig-7}
\end{figure*}

To adhere the conduction pathways of Na ions and the diffusion kinetics in B-NMFO cathode, the activation barrier and the conduction pathways of Na-ion are calculated through softBV-GUI-v131 software by using bond-valence-based empirical force fields \cite{ChenAC19, WongCM21}. The Na-migration barrier is found to be 0.155~eV between the two Na sites (Na1 and Na2), as shown by the vertical dashed line in Fig.~\ref{Fig-6}(a), which conclude that the Na ion migrates from Na2 site to Na1 site by surpassing the energy barrier of 0.155~eV via a 2D diffusion mechanism. In a simplified version, the Na$^{+}$ ion moves from one stable site (Na2) to another stable site (Na1) through the transition point barrier (saddle point, S1), which related to the hopping of Na ions in a specified path of Na2 $\rightarrow$ S1 $\rightarrow$ Na1. Here, each Na ion can hop to multiple adjacent sites, resulting in multi-directional 2D migration. In Fig.~\ref{Fig-6}(b), the 2D-migration pathways of Na ion seems to have honeycomb like network in the $ab-$plane and each Na1 is connected to three Na2 sites and vice-versa, activating trigonal hopping between the adjacent sites. Fig.~\ref{Fig-6}(c) represents the 3D view of conduction pathways symbolized by blue color iso-surfaces in between the octahedral slabs of MO$_{2}$ (yellow color) or inter-slab gallery. The  energy barrier (0.155~eV) between the sites and 2D conduction pathways in $ab-$plane are the governing factors for understanding the ion kinetics in these cathode materials. To further elucidate the effect of temperature on the Na-ion kinetics, distribution of relaxation times (DRT) analysis is performed on the EIS data. The DRT analysis is a mathematical tool for detailed analysis and diagnosis of different electrochemical processes occurring inside the cell at different relaxation times \cite{PatiJPS24}. This tool also attributes higher resolution than the conventional EIS representation through Nyquist or Bode plots. We use Kramers-Kronig (K-K) criterium to check the validity and authenticity of the impedance data using LIN-KK software, as described in Fig.~S4 of \cite{SI}. We can also get quantitative information regarding the reaction kinetics and contribution of individual physical processes to total cell resistance by using peak area, height and position of the DRT peaks. Therefore, the total resistance and relaxation time can be related as \cite{PatiJPS24}:
\begin{equation}
Z(\omega) = R_{0} + R_{p}\int_{0}^{\infty}\frac{g(\tau)}{1+j\omega\tau}d\tau
\end{equation}
where {\it R}$_{0}$ is the ohmic resistance (at high frequency intercept), {\it R}$_{p}$ is the polarization resistance (low frequency intercept), {\it g}($\tau$) is the distribution function of relaxation time in electrochemical processes which satisfy the relation as written below:
\begin{equation}
\int_{0}^{\infty}g(\tau)dln\tau = 1
\end{equation} 
Initially, the recorded Nyquist plots at different temperatures in Fig.~\ref{Fig-6}(d) are evaluated through the MATLAB program according to the Schmidt {\it et al.} study \cite{SchmidtJPS13} to garner the individual contributions at different relaxation times ({\it g}($\tau$) =  $\gamma(\tau)$). Figs.~\ref{Fig-6}(e--j) show the plots between impedance $\gamma(\tau)$ and logarithmic function of $\tau$, where $\tau$ is related to frequency through the equation $\tau$ = 1/2$\pi$$f$. The low frequency regime of the Nyquist plot with sharp slope originates the strongest peak P1 with a slow time constant of $\tau=$ 34~s, representing the mass transfer through diffusion of Na ions. This can also be visualized as the final binding sites of Na ions in the SEI layer and the bulk diffusion through the host. Two additional processes, P2 and P3, can be identified between $\tau$ = 10$^{-1}$~s and 10~s and one distinct and invariant peak P4 is found at $\tau$ =  10$^{-2}$ s. Here, the P2 and P3 are related to the charge transfer resistances of diffusive Na$^{+}$ through the cathode electrolyte interface (CEI) \cite{SoniESM22}. The interaction between processes P1 and P2 can be elaborated as follows: if P1 results from the diffusion of Na ions into the bulk material, it will remain useful throughout the battery cycling process. The conductivity of this bulk layer will enhance with both elevated Na ion concentration and increased driving potentials through hopping mode of charge transport, thereby attributing the minor shift and diminished relaxation strength with temperature. Nonetheless, the P1 peak remains present in the DRT, as the sodium ions must be present throughout the sodiation process. The P2 peak predominantly transpires during the low-slope phase, which can be attributed to the diffusive transfer of sodium ions into the host at high adsorption energy binding sites \cite{SoniESM22}. The faster process P3 is nearly invariant after 25\degree C and speculatively represent the faster electron transport in the defective sites. The last peak P4 visible at all temperatures and possess comparatively fast time constant   originating due to the electron transfer process at the metallic anode, which commonly known as Na plating \cite{SoniESM22}. 

\begin{figure*}
\centering
\includegraphics[width=\textwidth]{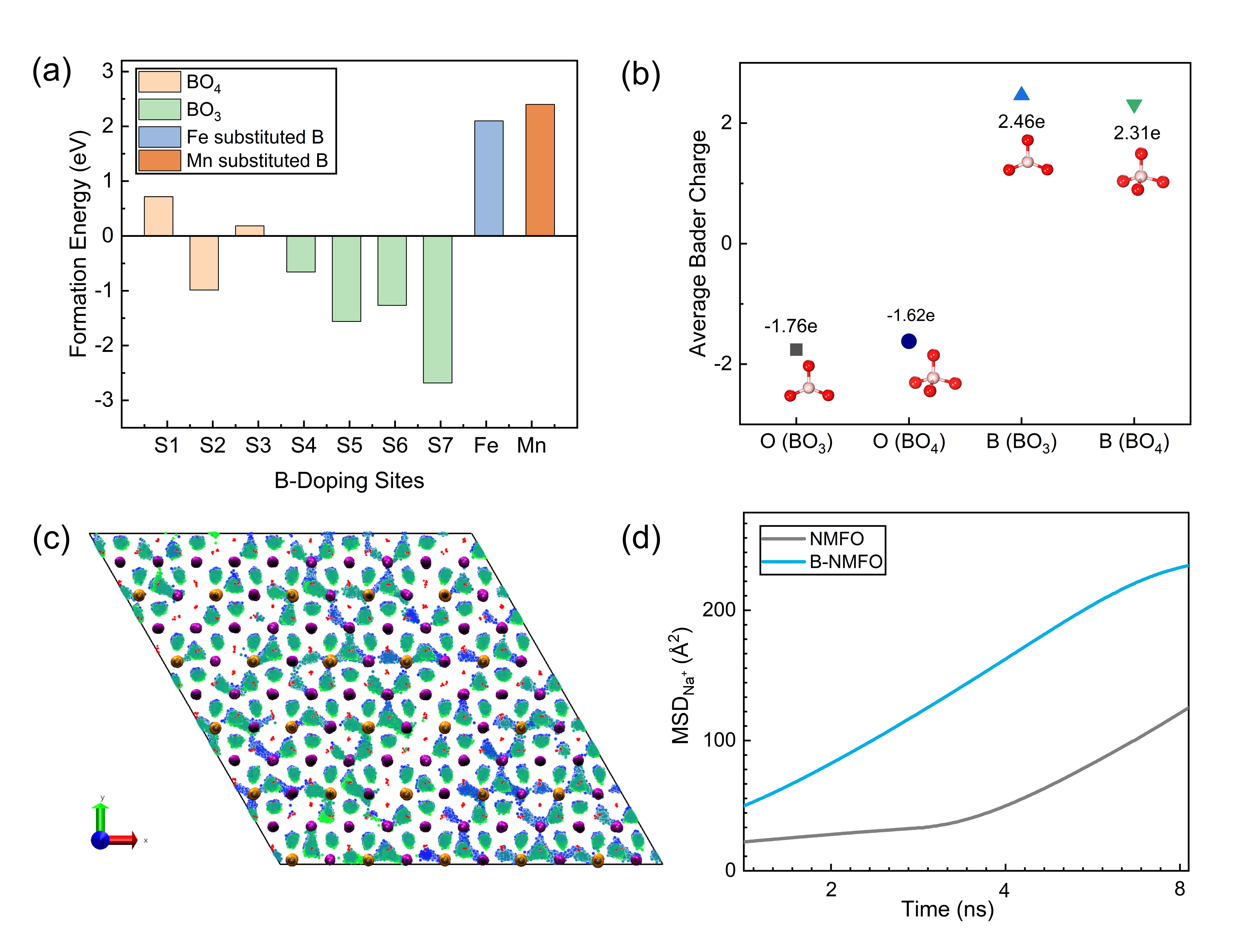} 
\caption {(a) The DFT calculated formation energies for B-NMFO system for possible boron occupancy at seven distinct sites for comparison. (b) The Bader charge distribution, (c) the Na-ion trajectory density maps (from green to blue) showing diffusion pathways in the bulk NMFO structure (color scheme: O--red, B--orange, Fe--black, Mn--purple). (d) The mean squared displacement (MSD) versus time plot for the Na-ion diffusion.}
\label{BNMFO-DFT}
\end{figure*}

We now investigate the oxidation states of individual elements and chemical composition through XPS measurements for both the samples where the binding energy scale is calibrated keeping the C 1$s$ peak at 284.6 eV and deconvolution of core-levels is done using Pseudo-Voigt function \cite{PatiJPS24}. The Mn~2$p$ deconvoluted core-level spectra in Figs.~\ref{Fig-7}(a, e) show spin-orbit splitting components 2p$_{3/2}$ and 2p$_{1/2}$ located around 642.1~eV and 644.1~eV for 2p$_{3/2}$ and 653.6~eV and 655.5~eV for 2p$_{1/2}$ which are associated to 3+ and 4+ states, respectively \cite{KimAMI20}. We also observe satellite peaks at 647.8~eV and 655.5~eV having splitting of 5.7~eV for both the samples \cite{Sharma_S_25}. In Figs.~\ref{Fig-7}(b, f), we show the Fe 2$p$ core-level spectra confirm 3+ oxidation state having 2p$_{3/2}$  and 2p$_{1/2}$ peaks at around 711.8~eV and 725.5~eV, respectively along with the corresponding satellite peaks 8.2~eV below the main peaks \cite{PatiJPS24}. Here, the Fe$^{3+}$ helps the structural stability by compensating the Jahn-Teller distortion arising due to the presence of Mn$^{3+}$ \cite{JiaJMCA22}. The authenticity of layered oxide is understood by an edge shared (Na$_{e}$) and face shared (Na$_{f}$) sites which are clearly located at 1071 and 1072.7~eV, respectively for both the NFMO and B-NFMO samples, see Figs.~\ref{Fig-7}(c g) \cite{IslamCEJ22}. Further, the O~1s core-level spectra in Figs.~\ref{Fig-7}(d, h) are deconvoluted in four components where the peak at 529.6~eV is associated to the lattice oxygen and the peak at 531.0~eV is the characteristic signature of oxygen vacancies, which are normally evolved in the oxide based systems due to the charge compensation \cite{IslamCEJ22}. Interestingly, here we observe relative increase in 531.0~eV peak intensity, whereas 529.6~eV peak intensity decreases, which indicate the increase of oxygen vacancies or a reduction in lattice oxygen intensity, respectively, upon boron doping \cite{SuAEM24, LuoAMI22, GuoNC21}. This in turn enhance Na$^+$ diffusion, modify the electronic structure, and facilitate the P2 $\rightarrow$ OP4 transition, which can lead to additional capacity contribution particularly around the ~3.5 V plateau \cite{GuoNC21}, as is consistent with discussion above. Also, the dQ/dV plots in Fig.~\ref{Fig-3}(h) showing features near ~3.5 V for the B-NMFO cathode support this interpretation. Further, the peaks positioned at 532.2 and 534.7 eV are attributed to the surface related oxygen species such as O-H and O-C, respectively \cite{KimAEM22}. Fig.~\ref{Fig-7}(i) depicts the B 1$s$ core-level spectrum with two deconvoluted peaks positioned at 192~eV and 193.7~eV elucidating the interstitial boron and the surface boron oxide \cite{RenEZ23, GuoESM24}.

Finally, we use density functional theory (DFT) to calculate the formation energies of boron (B) doping at seven symmetry-inequivalent tetrahedral interstitial sites (S1--S7), together with substitutional B at Fe and Mn sites, see Fig.~\ref{BNMFO-DFT}(a). The classification of S1--S7 follows the tetrahedral interstitial configurations defined in ref.~\cite{ChenJEC25}, where sites S1--S3 are located beneath occupied Na positions and typically retain BO$_{4}$ tetrahedral coordination, while the sites S4--S7 lie beneath Na vacancies and tend to relax into BO$_{3}$ trigonal planar configurations. Fig.~\ref{BNMFO-DFT}(a) demonstrates that there is a clear energetic difference between these two groups, i.e., sites S1--S3 (BO$_{4}$--type) have formation energies that are higher than average where S1 and S3 are slightly positive, while S2 is moderately negative. On the other hand, vacancy-adjacent sites S4--S7 have much lower and negative formation energies. This means that interstitial B incorporation is more likely to happen in Na-vacancy environments where the S7 site has the lowest formation energy (about --2.68~eV) making it the most stable configuration. The energetic stabilization of S4--S7 can be attributed to two main factors: (1) the reduced Na--B electrostatic repulsion in vacancy-adjacent sites, allowing B to relax downward, and (2) the formation of strong, short B--O covalent bonds (~1.35--1.45 \AA), which are characteristic of BO$_{3}$ units and contribute substantial bonding stabilization. The progressive decrease in formation energy from S4 to S7 further suggests that the local transition-metal (TM) environment plays a secondary but important role. In particular, the S7 site corresponds to a Mn-rich octahedral framework, where the surrounding MnO$_{6}$ units accommodate local lattice distortion more effectively, thereby minimizing the strain energy. This means the interstitial tetrahedral sites, especially those next to vacancies, are the dominant incorporation pathways for B.

Interestingly, these results show that B doping is very site-selective, with thermodynamic driving force favoring BO$_{3}$--type configurations under Na vacancies. The strong stability of S7 suggests that engineering Na vacancies could be a feasible way to control how much boron gets into Na-layered cathodes and improve their structural stability. To elucidate the origin of the different structural stabilities between BO$_{3}$ and BO$_{4}$ configurations, Bader charge analysis is performed, and the averaged charges of B and its coordinated O atoms are depicted in Fig.~\ref{BNMFO-DFT}(b). The B--O bond strength is greater in BO$_{3}$ than BO$_{4}$ due to the significant charge transfer from B to coordinating O, which is observed from the Bader charge on B in BO$_{3}$ is +2.46e versus +2.31e in BO$_{4}$, while the coordinated O atoms in BO$_{3}$ have a more negative electronic charge (-1.76e) than in BO$_{4}$ (-1.62e). The charge redistribution between B and O reduces to minimal charge transfer in BO$_{4}$ compared to BO$_{3}$. Additionally, this enhanced bond strength is attributed to the fact that the absence of electrostatic repulsion from Na allows the B atom, in the vacancy-adjacent coordination, to relax downwards forming a planar BO$_{3}$ unit. In contrast, when located beneath Na, electrostatic repulsion pushes B upward, forcing the formation of a BO$_{4}$ tetrahedral coordination with slightly weakened individual B--O bond strength. Importantly, the more positive B charge in BO$_{3}$ suggests a stronger electron-donor role of B, which enhances orbital hybridization between B 2$p$ and O 2$p$ states. Therefore, the larger charge separation observed in BO$_{3}$ (higher B positive charge and more negative O charge) confirms that the planar BO$_{3}$ configuration is electronically more stabilized than BO$_{4}$ in the lattice. This charge redistribution provides a fundamental electronic explanation for the lower formation energies and superior structural stability observed for vacancy-associated B doping sites.

Further, the MD simulations using interatomic pair-potential models are employed for understanding transport of Na-ion in NMFO and B-NMFO structures. The suitability of the potential parameters is initially assessed based on the cell parameter changes. As listed in Table S2 of \cite{SI}, only a minor change ($<$ 1\%) in lattice parameters of NMFO structures is observed, indicating a stable configuration with the applied potential model. This pair-potential model is also successfully applied to the similar layered oxide battery materials to understand the dynamics of Na-ions \cite{SethAEM25, DengJNCS16, KieuJNCS11}. Fig.~\ref{BNMFO-DFT}(c) illustrates the density map of Na-ions, showing the diffusion pathways indicated by the color changes from green to blue across the MD trajectories. Also, Fig.~\ref{BNMFO-DFT}(d) shows the MSD versus time plot, which show higher values for the B-NMFO sample. The Na-ion self-diffusion coefficient ({\it D$_{Na^{+}}$}) is estimated to be ~7.22 (± 0.86)$\times$10$^{-10}$ cm$^{2}$s$^{-1}$ for the NMFO system and ~1.37 (± 0.36)$\times$10$^{-9}$ cm$^{2}$s$^{-1}$ for the B-NMFO system, which are consistent with GITT analysis discussed above. 

\section{Conclusions}

In summary, this work elucidates the role of boron doping in modulating the electrochemical behavior and diffusion kinetics of layered NMFO cathodes for sodium-ion batteries. The improved performance of the B-NMFO cathode, delivering a high discharge capacity of 163 mAhg$^{-1}$ at 0.1 C and ~70\% capacity retention at 1 C after 200 cycles (as compared to 133 mAhg$^{-1}$ and 60\% for the NMFO), is attributed to a synergistic interplay of structural and electronic effects induced by boron incorporation. The analysis of CV and GITT reveal the pseudocapacitive nature of capacity and diffusion coefficient values in the range of 10$^{-8}$--10$^{-10}$ cm$^{2}$s$^{-1}$ for both the cathodes. The boron doping modifies the local oxygen environment and introduces controlled oxygen non-stoichiometry, which enhances the Na$^+$ diffusion kinetics and alters the electronic structure of the host lattice. Further, the temperature dependent DRT and migration pathway analysis elucidates the individual physical reactions at a particular time and energy barrier for Na-ion migration at different sites, respectively. The DFT calculations suggest B doping in NMFO structure is site-selective, favouring BO$_{3}$-type configurations under Na-ion vacancies as indicated by the lower energy of formation. The changes facilitate a more pronounced and reversible P2 $\rightarrow$ OP4 phase transition at ~3.5 V, providing additional high-voltage capacity.  Further, the Bader charge analysis verifies that the in-plane BO$_{3}$ configuration is geometrically favourable through stronger B-O covalent interactions. The MD simulations further reveal that the B-NMFO structure shows higher Na-ion diffusivity as compared to the NMFO.  

\section{\noindent ~Acknowledgments}

JP and PS thank the UGC and DST, respectively for the fellowship support. We acknowledge the DST for financially supporting the research facilities for sodium-ion batteries through {\textquotedblleft}DST-IIT Delhi Energy Storage Platform on Batteries" (project no. DST/TMD/MECSP/2K17/07) and from SERB-DST (now ANRF) through a core research grant (file no.: CRG/2020/003436).  We thank IIT Delhi for providing research facilities for sample characterization (the XRD and Raman at the physics department; as well as the FE-SEM, EDS, and XPS at CRF). DS and MAH acknowledge the high-performance computing (HPC) facility at IIT Delhi for providing the computational resources used. 

\section{\noindent ~Conflict of interest}

The authors declare that there are no conflicts of interest associated with this article. 

\section{\noindent ~Data Availability}

The data that support the findings of this study are available upon reasonable request.

\end{document}